\documentclass{elsart}
\usepackage{amssymb}
\usepackage[dvips,unicode]{hyperref}

\usepackage[dvips]{graphicx}

\begin{document}

\begin{frontmatter}

\title{Theoretical study of cache systems}

\author{Dmitry Dolgikh}

\address{Samara State Aerospace University, Moscovskoe sh. 34a,
Samara, 443086, Russia}

\author{Andrei Sukhov\corauthref{avt} }
\address{Laboratory of Network Technologies, Samara Academy
of Transport Engineering, 1 Bezymyanny per., 18, Samara, 443066,
Russia}

\thanks[avt]{Corresponding author\\{\em E-mail addresses:
sukhov@ssau.ru}~(Andrei M. Sukhov), {\em ddolgikh@ssau.ru}~(Dmitry
G. Dolgikh)}

\begin{abstract}

The aim of this paper is a theoretical study of a cache system in
order to optimize proxy cache systems and to modernize
construction principles including prefetching schemes. Two types
of correlations, Zipf-like distribution and normalizing
conditions, play a role of the fundamental laws. A corresponding
system of equations allows to describe the basic effects like
ratio between construction parts, steady-state performance,
optimal size, long-term prefetching, etc. A modification of the
fundamental laws leads to the description of new effects of
documents' renewal in the global network. An internet traffic
caching system based on Zipf-like distribution (ZBS) is invented.
The additional module to the cache construction gives an effective
prefetching by lifetime.

\end{abstract}

\end{frontmatter}

\section{Inroduction}
\label{Intr}

Rapid development of computer technologies in the end of the
last millennium leads to appearance of a virtual world with its
own laws. Unfortunately, in this period too few attention was
given to studying of fundamental principles of virtual
life.

The main constructions of the virtual world base on the concrete
algorithms intended for providing vital functions. The
easiest and obvious principles were used for constructing and
modernizing of these algorithms. During their implementation weren't
any time for studying and optimization. Often their low performance
was compensated with growing power of computers. The main thing was
that the new constructions of the virtual world allowed to manage
the raising problems.

Now days we can afford some respite and part of forces should
be transferred to studying of fundamental laws and to
optimization of vital systems on a base of recent
knowledge. Frankly speaking, any research area may be considered
as scientific field if and only if its basic principles are
enveloped in a mathematical form and new rules may be
predicted on a base of known facts.

The aim of this paper is a theoretical study of a cache system in
order to find a way for optimization of proxy systems and to
modernize construction principles. Principles that should be used
for a model is discussing as well. A universal model can be easy
generalized to any applications based on Zipf-like distribution
like prefetcnig schemes, Content distribution Networks
(CDN)~\cite{gadcr}, peer-to-peer systems~\cite{rfi,srip}, Internet
search engine, etc.

The rest of the paper is organized as follows.
Section~\ref{parameters} provides some background information
about model parameters. Section~\ref{anal-laws} presents the
analytical laws and the special points of theoretical model.
Section~\ref{constr} describes the main elements of cache
construction. In Section~\ref{steady} we examine steady-state or
limiting performance of static caching schemes when Web documents
were not changing and new documents were not being generated.
Section~\ref{optim} discusses the ways of optimization of cache
systems. New principles of cache construction are formulated in
Section~\ref{const1}. Section~\ref{renew} discusses the
mathematical description of renewal of Web documents. The benefit
of web prefetching is the contest of Section~\ref{prefet}.

\section{Model parameters}
\label{parameters}

In this section definitions of variables are given and
basic approaches and terminus has been analyzed.

\begin{figure}
\begin{center}
\includegraphics[width=0.35\textwidth]{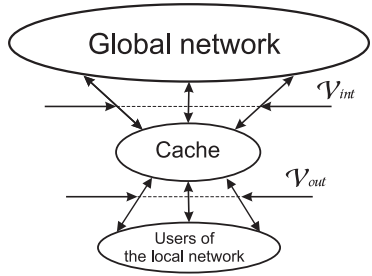}
\end{center}
\caption{Scheme of the proxy cache} \label{scheme-e}
\end{figure}

In today's common web configurations, the proxy server exists
between clients and web servers. Clients send all requests to a
global network as it is shown on the Fig.~\ref{scheme-e}. Some
documents are requested several times and therefore they should be
held in a cache system.

As it was reported in Refs.~\cite{alm1,bres,nish} the relative
frequency of requests to Web pages follows Zipf-like
distribution~\cite{zipf}. This distribution states that the
relative probability of a request for the $i$'th most popular page
is
\begin{equation}
\label{zipf-gen}
  p_i=\frac{A}{i^\alpha},
\end{equation}
where $A=p_1$ is the probability of the most popular item and
$\alpha$ is positive exponential value less then unity.

Let the users ask for $K$ documents during time $t$:
\begin{equation}
\label{k-doc} K(\nu_{out},t)=C\nu_{out}t.
\end{equation}
This result depends on aggregated bandwidth $\nu_{out}$, on time
$t$ and a constant $C$. The constant $C$ from Eq.~(\ref{k-doc}) is
an inverse proportion of a mean size $E(C)$ of documents
received from the global network:
\begin{equation}\label{meanC}
  C=\frac{1}{E(C)}
\end{equation}

In Ref.~\cite{wolm} a number of received documents $K$ is
defined by request stream from a population of $N$ users
\begin{equation}
\label{k-user} K(\lambda,N,t)=\lambda Nt,
\end{equation}
where $\lambda$ is average client request rate.

\begin{figure}
\begin{center}
\includegraphics[width=0.35\textwidth]{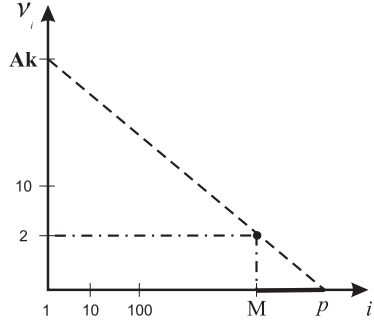}
\end{center}
  \caption{Special points of cache system}
  \label{M-p}
\end{figure}

Let $p$ documents be unique and $M$ be equal to quantity of
documents that can be requested from the cache system as it has
shown on the Fig.~\ref{M-p}. Then an efficiency of the system or a hit
ratio can be defined as
\begin{equation}
\label{eff-1} H=\frac{k-p}K= \int\limits_1^M\frac{A}{x^\alpha} dx
\end{equation}
the ratio of non-unique cacheable documents $k-p= k
\int_1^M\frac{A}{x^\alpha}dx$ to their total number $K$.

Wolman et al.~\cite{wolm} examined steady-state or limiting
performance of caching schemes. Their model of steady-state
performance assumes that a cache can store all cacheable documents
in the Web and there isn't any capacity misses in workload. The
probability that a requested document is cacheable is $p_c$,
($k=p_cK$). With this assumptions the maximal rates of $H$ is
identical with $p_c$ , i.e. the ideal hit rate for cacheable
documents $H_i=H/p_c$ would approach $100\%$.

A key difference between the models presented by Breslau at
al~\cite{bres} and by Wolman et al~\cite{wolm} is that document
rate of change has been incorporated into the model rather than
assuming that documents are static. In order to describe this
effect an additional multiplier has been included to a integral
from Eq.~(\ref{eff-1}):
\begin{equation}\label{add-m}
\lambda Np_i/(\lambda Np_i+\mu),
\end{equation}
where $\mu$ is an exponential parameter of interarrival times for
object requests and updates, $p_i$ is proportional to
$1/i^\alpha$. This multiplier throws off one effective query of
any document from cache during its lifetime $T_{ch}=1/\mu$, that
is the interval between two consecutive modifications of object.
Such an updating request must be redirected to the global network
to get the renewed Web page.

\section{The analytical laws and the special points}
\label{anal-laws}

An analytical model is constructed with a goal to predict behaviour
of the investigated systems on the base of the analytical laws
written in a mathematical form and to find ways for optimization.
Mathematical constructions of computer models remind laws
of the thermodynamics and the molecular physics in many respects.
There are two directions of modeling: macroscopic
and microscopic approaches.

In the first case the corresponding laws are formulated for parameters
describing cache systems as an indivisible object. The following
values should be considered as macroscopic parameters: hit ratio
$H$, Zipf exponent $\alpha$, the number of unique documents $p$,
$M$, etc.

The microscopic description assumes an analysis on a level of
transmitting IP packets, of requests to the global network, i.e.
huge number of the events. In this case investigators have
to apply the theory of stochastic processes such as Poisson's
distribution or Markov's chain. The main problem of this approach
is a right interpretation of analytical results and their
generalization, i.e in a limit passage from microscopic values to
macrolaws.

Usually the macroscopic dependencies operate with values that
are measured directly. Unfortunately, corresponding laws are not
yet found for the main part of the virtual world therefore
a particular attention should be given to their search. Proxy
cache is a pleasant exception to the general rule.

Two types of correlations pretend to a role of fundamental
laws described cache systems~\cite{dols}:
\begin{itemize}
  \item A Zipf-like distribution, see Eq.~(\ref{zipf-gen})
  \item Normalizing conditions or a sum of the
  probability to request the universe of $1 \leq n \leq k$ objects.
\end{itemize}

Mentioned laws could be applied to the special points of Zipf
distribution. As it is shown on the Fig.~\ref{M-p} there are a few
special points but only two of them, $M$ and $p$, are used for
construction of the theory. Then the Zipf-like distribution leads
to
\begin{eqnarray}
\label{A_M} \frac{Ak}{M^\alpha} &= & 2,\\ \label{A_p}
\frac{Ak}{p^\alpha} & =& 1.
\end{eqnarray}
Normalizing conditions for the first $M$ and $p$ documents from a
cache are
\begin{eqnarray}
\label{Hi-n} \int\limits_1^M \frac{A}{x^\alpha} dx &=& H_i,\\
\label{1-n} \int\limits^p_1\frac{A}{x^\alpha} dx &=& 1.
\end{eqnarray}
Here $H_i$ is an ideal (steady-state) performance that assumes
unlimited capacity. For a real system the Eq.~(\ref{Hi-n}) has
been transformed to
\begin{equation}\label{H-r}
 H=p_c\int\limits_1^{S_k}\frac{A}{x^\alpha}dx,
\end{equation}
where $S_k$ is a number of cache objects requested repeatedly,
i.e. $\vartheta_i\geq 2$ ($\vartheta_i=p_i k$) for $i\leq S_k$.

Five Egs.~(\ref{A_M}-\ref{H-r}) allow to describe the basic
effects, to predict parameters value, ways of optimization, etc.
Modification of this system of five equations would lead to
describing of new effects.

\section{Elements of cache construction}
\label{constr}

This section discusses which elements are compulsory for
cache system. Also possible ratios between these elements
should be investigated.

Three main parts are distinguished in any cache system:
\begin{itemize}
\item
A kernel $S_k$ that contains popular documents with $\vartheta_i\geq
2$.
\item
An accessory part $S_u$ that keeps unpopular documents requested from
the Internet once, i.e. $\vartheta_i=1$.
\item
A managing part $S_m$ that contains statistics of
requests and rules for replacement of cache objects.
\end{itemize}

A system of the Eqs.~(\ref{A_M}-\ref{H-r}) determines mathematical
correlations between cache elements $S_k, S_u, S_m$.

First of all it is necessary to write the expressions for
experimental search of $\alpha$ With help of the
Eqs.~(\ref{A_M},\ref{A_p})
\begin{equation}\label{S-k-e}
  p=2^{1/\alpha_1}M.
\end{equation}
The correlation between $p$ and $k$ from system of
Eqs.~(\ref{A_p},\ref{1-n}) gives
\begin{equation}\label{c-kp}
  k=\frac p {1-\alpha_2}.
\end{equation}
Another useful relation is a solution of a system of
Eqs.~(\ref{A_M},\ref{Hi-n}) concerning variable $M$:
\begin{equation}\label{M-ker}
  M=\frac{(1-\alpha_3)H}{2}K.
\end{equation}
This means that only each 15th document should be stored in the
cache among all documents requested from the global network (for
typical values $\alpha\geq 0.7, H\geq 35\%$).

\begin{figure}
\begin{center}
\includegraphics[width=0.45\textwidth]{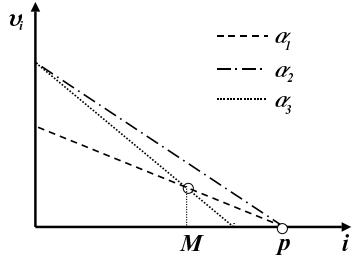}
\end{center}
  \caption{Possible $\alpha$ values}
  \label{Alp}
\end{figure}

It should be noted that three Eqs.~(\ref{S-k-e}-\ref{M-ker}) give
the different values of $\alpha$
\begin{equation}\label{dif-al}
  \alpha_1 < \alpha_2 < \alpha_3.
\end{equation}
These differences $\alpha_3-\alpha_2$, $\alpha_2-\alpha_1$ are of
order $0.04\div 0.05$ and may be graphically explained, see
Fig.~(\ref{Alp}). As a rule the value $\alpha_3$ from
Eq.~(\ref{M-ker}) is found from experimental data and it is used
for the further calculations.

Analyzing {\em log\/}~files of proxy cache a mean lifetime $t_u$
can be calculated for those documents, which popularity is
$\vartheta_i=1$. Such a statistic determines also lifetime
$T_{eff}$ of cache objects with a citing index $\vartheta_i=2$,
i.e. those items that have been stored in proxy cache, requested
one time from a proxy and have been deleted subsequently.

It is easy to see that a number of documents in a cache kernel
$S_k$ is a value $M(T_{eff})$ corresponding to the time $T_{eff}$,
and $S_u$ depends on $p, M, t_u$:
\begin{eqnarray} \label{S-k-n} S_k &= &
M(T_{eff}),\\ \label{S-u-n} S_u &= & p(t_u)-M(t_u),
\end{eqnarray}
The following ratio between the part $S_k, S_u$ has been found
with help of Eqs.~(\ref{k-doc}),~(\ref{S-k-e}):
\begin{equation}\label{sk/su-n}
 \frac{S_k}{S_u}=\frac{T_{eff}}{(2^{1/\alpha}-1)t_u}
\end{equation}

The Eq.~(\ref{sk/su-n}) gives a criteria for optimal using of a
storage capacity. This criteria allows to plane an experiment: how
does efficiently of cache replacement algorithm depend on
correlation of the proxy parts? It should be noted this result as
well as the experiment plan was the consequence of the theoretical
study.

The analysis of {\em log\/}~files from Samara State Aerospace
University proxy~\cite{dols2} leads to the facts that variables
$t_u$ and $T_{eff}$ are directly proportional to the cache size
$S_{eff}/\nu_{int}$ and can be considered as coincided values:
\begin{equation}
\label{t=T} T_{eff}\simeq t_u.
\end{equation}

In other words the kernel and accessory parts are approximately
correlated as 1:2 or less then 40\% of storage capacity has been
used for the basic goal to store the repeatedly requested
documents.

\section{Steady-state performance (static Web)}
\label{steady}

In this section we examine steady-state or limiting performance of
static caching schemes when Web documents were not changing and
new documents were not being generated. Wolman et al
model~\cite{wolm} of steady-state performance assumes that a cache
can store all cacheable documents in the Web and there aren't any
capacity misses in the workload. With this assumption, they
conceive that in the long term the hit rate $H_i$ for cacheable
documents would approach 100\%. However, considerable part of
cacheable documents $p-M$ is requested from the global network
only one time as it is shown on Fig.~\ref{M-p} that illustrate
Zipf-like distribution.

More accurate expression for ideal hit ratio $H_i$ may be written
with help of Eqs.(\ref{S-k-e},\ref{c-kp},\ref{M-ker}) to in
consecutive order $k,p,M$
\begin{equation}\label{H-i-a}
  H_i\leq 2^{(\alpha-1)/\alpha},
\end{equation}
that gives 75\% for $\alpha=0.7$.

The alternative expression for ideal hit ratio $H_i$ follows from
Fig.~\ref{M-p}:
\begin{equation}\label{H-i}
  H_i \leq 1-\frac{p-M}{k}
\end{equation}
that gives 80\% \cite{dols2}.

The hit rate $H$ of web cache is considered to grow in a log-like
fashion as a function of cache size \cite{alm,bres,caoi}. Wolman`s
work \cite{wolm} shows that there is no benefit in additional
cache size beyond a certain size. However such an statements are
impeccable only for an extremely huge sizes.

The cache administrators believe that a production cache should
store about 2-3 days of traffic. Such an recommendation can be
found on the cite of the Measurement Factory Inc., that tested
well-known cache products~\cite{meas-f}.

From analyzes of the Eq.~(\ref{H-r}) for cache performance the
following dependence appears
\begin{equation}\label{H-S}
  \frac{H_1}{H_2}=\left( \frac{S_1}{S_2}\right)^{1-\alpha},
\end{equation}
that allows to talk about power fashion \cite{dols2} in size's
area of a few days traffic. This fact has been confirmed by
experimental research~\cite{dols2}.

\section{Optimization of cache systems}
\label{optim}

A fine tuning of cache that increases hit ratios by only several
percentage points would be equivalent to a several-fold increase
in cache size. In order to determine an optimal cache size Kelly
and Reeves~\cite{kelr} are guided on economical methods like
monetary cost of memory and bandwidth.

The aim of our theoretical investigation is a search of
dependencies between cache size $S=S_k+S_u$ and other
macroscopic parameters like internal bandwidth $\nu_{int}$,
maximum hit ratio $H_i$, Zipf's exponent $\alpha$, etc.

As it is shown in Ref.~\cite{dols} a ratio of effective size of
cache system $S_{eff}$ to aggregated bandwidth of external
links $\nu_{int}$ can be considered as a constant:
\begin{equation}\label{base-r}
  \frac{S_{eff}}{\nu_{int}}=\tau.
\end{equation}
Our goal is to calculate a lower limit $\tau(\mu,\alpha)$ when the
performance mounts to the 35\% level $(H_{eff}\geq 35\%)$ of well
working system.

Eqs.(\ref{H-i-a},\ref{H-i}) give the following estimation for a
real system with $\alpha\geq 0.7$, $p_c=0.6$:
\begin{equation}\label{eff-H}
\frac{p_c H_i}{\sqrt{2}}\ge H_{eff}\ge 35\%
\end{equation}

Eqs.(\ref{Hi-n},\ref{H-r}) allow to calculate the correlation
between documents number in a real cache kernel $S_k$ and a
number $M_{max}$ corresponding to limiting performance $H_i$:
\begin{equation}\label{eff-M}
 S_k= 2^{1/(2(\alpha-1))} M_{max}
\end{equation}
where
\begin{equation}\label{M-max}
  M_{max}=\frac{(1-\alpha)p_cH_i}{2} \nu_{out}T_{ch}
\end{equation}
Parameter $T_{ch}=1/\mu_u=186$ days is the time between renewal of
unpopular documents~\cite{wolm}. If $H_{eff}=0.35$, then
\begin{equation}\label{tau-f}
\tau(\mu_u,\alpha)=2^{1/\alpha} 2^{1/(2(\alpha-1))}
\frac{1-\alpha}{2.61\mu_u}.
\end{equation}
It is easy to calculate $\tau=6,0$ days for $\alpha=0.8$.

\section{New principles of cache construction}
\label{const1}

The main result of any research of cache system should be finding
a way for increasing hit ratio. We see the two basic directions
for achievement that goal. The first conjecture is that the hit
rate $H$ increases with growth of $S_k/S_u$. The second one is
based on the general feature of the existing cache algorithms that
the only documents requested two times and more during the time
$t_u$ are included in the kernel $S_k$. Therefore we conclude that
caching algorithm based on the rigid tie of the main cache
parameters to the time $t_u$ is low effectiveness. Our approach is
to separate the management block based on the requests' statistics
from the time $t_u$.

The current situation can be resolved with the new construction
principles for cache systems. The most important feature is that
fixed ratio between cache parts must be guaranteed. The statistics
of requests must be kept long time for all cacheable documents
including delayed ones. An inseparable part of the new
construction is a replacement algorithm with a metric function
based on Zipf law.

The basic principles of system of internet traffic caching based
on a Zipf-like distribution (ZBS) could be formulated as follows:
\begin{enumerate}
  \item
Three main construction parts are distinguished in ZBS-system:
\begin{itemize}
\item
A kernel $S_k$ that contains popular documents with
$\vartheta_i\geq 2$.
\item
An accessory part $S_u$ keeps unpopular documents requested from
the Internet once, i.e. with $\vartheta_i=1$. Size of the
accessory part $S_u$ must not exceed 10\% from the full size of
the ZBS cache $S_{eff}$.
\item
A managing part $S_m$ contains statistics of requests as a base
for a replacement policy. The control information including
statistics are kept in $S_m$ for a long time interval $t_s\geq
10(S_{eff}/\nu_{int})$, where $\nu_{int}$ is an aggregated
bandwidth of the external links. The value $t_s$ must exceed one
month, the maximal time interval is restricted by a half of an
year. A document popularity $\vartheta_i$ takes into account all
requests that made by users during the time $t_s$.
\end{itemize}

\item
An additional necessity to request a cached document from the
global network is caused by document modification. The newly
received document' item is stored in the kernel $S_k$ and the
corresponding frequency parameter assumes to be equals unity
$\vartheta_i=1$.

\item
An replicate metric $C_i=T^i_z/\vartheta_i$ is calculated for each
document stored in the kernel $S_k$, where $T^i_z$ is a time
period spent from the last modification of the document and
$\vartheta_i$ is the corresponding request's frequency. The
Fig.~\ref{TNe} contains a graphical explanation of these
variables. A document with the biggest $C_i$ is deleted when the
kernel $S_k$ is overfilled.

\item
A metric $C_i=\frac{T^i_z}{\nu_i}\frac{1}{E_i}$ is applied for
increasing byte hit ratio. Here $E_i$ is the size of the
corresponding document.

\item
The document from the accessory part $S_u$ is deleted

\begin{itemize}
  \item
when it is placed to the kernel $S_k$ after the second request,
  \item
accordingly recently-based policies, e.g. FIFO, when the accessory
part $S_u$ is overfilled.
\end{itemize}

\begin{figure*}
\begin{center}
\includegraphics[width=0.85\textwidth]{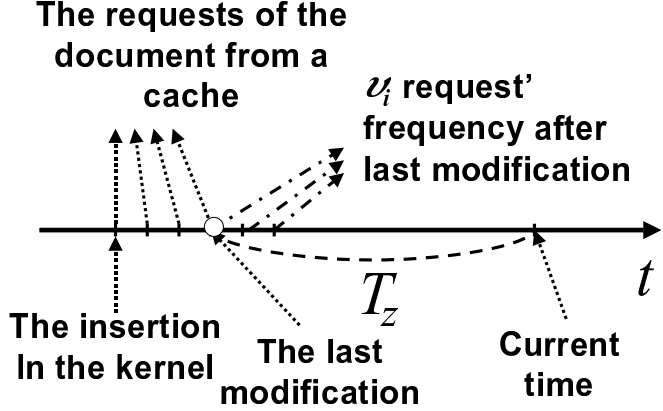}
\end{center}
  \caption{The illustration of variables included in metric function}
\label{TNe}
\end{figure*}

\end{enumerate}

The described cache scheme improves the hit ratio by a few percent
according Eq.~(\ref{H-S}) because of the kernel size $S_k$
increases twice from 40\% to 90\% when other things being equal.
For the huge cache size the profit consists of in an economy of
the half of disk storage.

\section{Renewal of Web documents}
\label{renew}

In the previous section our analysis has been based on the model
presented by Breslay et al~\cite{bres} and extended by Wolman et
al.~\cite{wolm} to incorporate document rate of change. The Wolman
model yields formulas to predict steady-state properties of Web
caching systems parameterized by population size $N$, population
request rates $\lambda$, document rate of change $\mu$, size of
object universe $n$ and a popularity distribution for objects.

The key formulas from Wolman et al.
\begin{eqnarray}
C_N
&=&\int\limits_1^n\frac{1}{Cx^\alpha}\left(\frac{1}{1+\frac{\mu
Cx^\alpha}{\lambda N}}\right) dx \label{C_N}\\ C
&=&\int\limits_1^{n}\frac{1}{x^{\alpha}} dx \label{C-old}
\end{eqnarray}
yield $C_N$, the aggregate object hit ratio considering only
cacheable objects. Document rates of change $\mu$ were considered
to take two different values, one for popular documents $\mu_p$
and another for unpopular documents $\mu_u$. An additional
multiplier from Eq.~(\ref{C_N}) throws off one effective query to
any document from cache during time $T_{ch}=1/\mu$ between its
changes. Such an request updates the existing objects when it has
been modified.

\begin{figure}
\begin{center}
\includegraphics[width=0.45\textwidth]{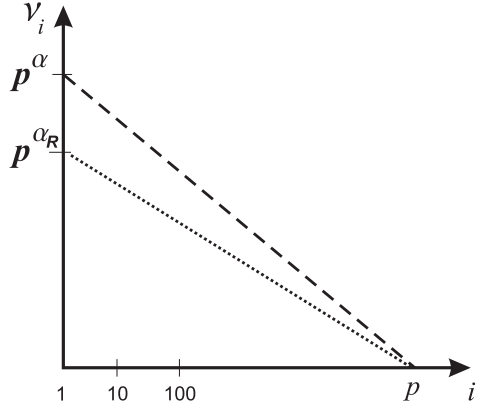}
\end{center}
  \caption{The renewal effect}
  \label{RZipf}
\end{figure}

In present paper we assume that the document rate of change
$\mu(i)$ depends on its popularity $i$. Mathematical equivalent of
this assertion is that the steady-state process of demand cache is
again described by Zipf distribution with a small $\alpha_R$ as it
is shown on Fig.~\ref{RZipf}.

The difference between an ideal performance of a cache system and
its real value has been conditioned by the renewal of documents in
the global network
\begin{equation}\label{deltaH}
  \triangle H=\left(
  \sum\limits_{i=1}^M \vartheta_i-HK\right)/K
\end{equation}

 The ideal Zipf-like distribution corresponds to the upper
line on the Fig.~\ref{RZipf} but the real hit ratio $H$ determines
$\alpha_R$ as
\begin{equation}\label{alR}
  \alpha_R=1-2M/HK
\end{equation}

Now it is easy to find $\mu_i$
\begin{equation}\label{mui}
  \mu(i)=\frac{\vartheta^{id}_i-\vartheta^R_i}{T_{st}}=
  \frac{(p/i)^\alpha-(p/i)^{\alpha_R}}{T_{st}}
\end{equation}
It should be noted again that the experiment's plan is the
consequence of the theoretical study. The corresponding results
have been calculated from the proxy' trace of Samara State
Aerospace University \cite{dols2} and has been assembled in the
Table~\ref{al-R-e}.

\begin{table*}
\caption{The renewal parameters}\label{al-R-e}{\scriptsize
\begin{tabular}{|c|c|c|c|c|c|}
\hline $\alpha$ & $\alpha_R$ & $\triangle H $ & $H$& $\mu_p$ &
$\mu_u$ \\ \hline 0.72 & 0.7 & 2.3\% & 32.04\% & 1/6.2 days &
1/202 days
\\ \hline
\end{tabular}}
\end{table*}

Finally, the system of Eqs.~(\ref{A_M}-\ref{H-r}) must be
modernized by change of variables $\alpha$ on $\alpha_R$.

\section{Long-term Prefetching}
\label{prefet}

One way to further increase the cache hit ratio is to anticipate
future requests and prefetch these objects into a local cache. The
benefit of web prefetching is to provide low retrieval latency for
users, which can be explained as high hit ratio.

This section examines the costs and potential benefits of
long-term prefetching for content distribution. In traditional
short-term prefetching, caches use recent access history to
predict and prefetch objects likely to be referenced in the near
future. In contrast, long-term prefetching uses long-term
steady-state object access rates and update frequencies to
identify objects to replicate to content distribution locations.

Using analytic models and trace-based simulations, Venkataramani
at al. \cite{venk}, Jiang at al. \cite{jiang} examine algorithms
for selecting objects for long-term prefetching. They use
threshold-based algorithms and fetch those objects with values
that exceed the threshold. The object selection criterion of
prefetching determines which object to replicate in advance. They
have several options of criteria such as object popularity and
lifetime.

An algorithm of prefetching by good-fetch was proposed by
Venkataramani at al.~\cite{venk}. As for object $i$, assume the
object's lifetime $l_i$, its probability of being accessed $p_i$,
and user request arrival rate denoting how many requests arrive
per second is $a$. Then the probability of object $i$ to be
accessed before it dies is
\begin{equation}\label{GD-f}
  P_{goodFetch}=1-(1-p_i)^{a\times l_i}
\end{equation}
where $a\times l_i$ is the number of requests arriving during the
lifetime of object $i$.

Jiang at al. \cite{jiang} has proposed an algorithm using the $a
p_i l_i$ value to select objects. Objects with the highest value
of $a p_i l_i$ will be included in the prefetching set, meaning
those objects with most possible requests will be prefetched in
caches.

The analytical model derived in~\cite{venk} allows to express the
hit ratio of threshold-based algorithms. In order to describe this
effect an additional multiplier has been included to a integral
from Eq.~(\ref{eff-1}). Such an fraction
\begin{equation}\label{ven-mul}
ff(i)= a p_i l_1/ (a p_i l_i + 1)
\end{equation}
is identical to Wolman multiplier from Eq.~(\ref{add-m}). The
fraction (\ref{ven-mul}) represents the hit ratio among accesses
to the object $i$. Stated otherwise, this denotes the probability
of object $i$ being fresh when it is being accessed. It is called
a freshness factor of object $i$ denoted as $ff(i)$ in
\cite{venk}.

The approach developed in the Section~\ref{renew} allows to
calculate freshness factor $ff(i)$ as coefficient between hit
ratios of demand cache and of ideal performance, see
Fig.~\ref{RZipf}. Taking into account the Eq.~(\ref{c-kp}) and the
data from the Table~\ref{al-R-e} it is easy to see
\begin{equation}\label{ffi}
  ff(i)=k_R/k_i=(1-\alpha_R)/(1-\alpha_i)= 0.93.
\end{equation}

For threshold-based algorithms, an object $i$ is always kept fresh
by prefetching it whenever there is any change, if its
corresponding metric is above its threshold value $T_p$. Then we
will always have a hit on object $i$, while the hit ratio for
other objects remains same as $ff(i)$. In other words the
prefetching algorithms allow to reach highest hit ratio
corresponding to upper line on Fig.~\ref{RZipf} to use an
additional bandwidth. This additional value has been estimated
\cite{venk}
\begin{equation}\label{add-ban}
  \Delta BW= (1-ff(i))\nu_{int}
\end{equation}
as 7\% to the mean magnitude of aggregated bandwidth of external
links from Eq.~(\ref{base-r}).

On our opinion the additional module to the cache construction
described in Section~\ref{const1} can give an effective
prefetching by lifetime:

\begin{itemize}
  \item
An additional mark $N^i$ is introduced for each cache item. It
equals the number of document modification for the time $T_{ins}$
from installation of cache system.

  \item
The threshold value is $T^i_p= T_{ins}/N^i$.

  \item
If the time from a last modification of the document $T_z^i$
exceeds corresponding threshold $T^i_p$ then the cache fetches
this object.
\end{itemize}

Mathematical basis of this module is uniform distribution of
modification moments inside $T_{ins}$.

\section{Conclusions}
\label{disc}

The aim of this paper is a theoretical study of a cache system in
order to optimize proxy cache systems and to modernize
construction principles including prefetching schemes. Two types
of correlations, Zipf-like distribution and normalizing
conditions, play a role of the fundamental laws. A corresponding
system of equations allows to describe the basic effects like
ratio between construction parts, steady-state performance,
optimal size, long-term prefetching, etc. A modification of the
fundamental laws leads to the description of new effects of
documents' renewal in the global network.

The main result of any research of cache systems is finding a way
for increasing hit ratio. Our theoretical study leads to the
criteria for an optimal using of storage capacity. This criteria
allows to plane several experiment on the base of the mathematical
analysis.

The current situation can be resolved with the new construction
principles for cache systems. An internet traffic caching system
based on Zipf-like distribution (ZBS) is invented. The new system
construction consists of the three parts: a kernel for storing
popular documents, an accessory part for unpopular documents and a
management part for keeping requests' statistic. Documents'
replacement algorithm based on a metric function that uses
Zipf-like distribution is a part of the new ZBS system. The
additional module to the cache construction described in
Section~\ref{const1} gives an effective prefetching by lifetime.

\end{document}